\documentclass[
 ,final            % use final for the camera ready runs
  ]
  {aipproc}

\layoutstyle{6x9}

\usepackage{epsfig}
\usepackage{graphicx}
\usepackage{bm}
\usepackage{amsmath}
\usepackage{amssymb}

\newcommand{\si}{\sigma}

\newcommand{\az}{\varphi}
\newcommand{\ro}{\rho}

\newcommand{\oeq}{\begin{equation}}
\newcommand{\ceq}{\end{equation}}
\newcommand{\oeqn}{\begin{eqnarray}}
\newcommand{\ceqn}{\end{eqnarray}}

\renewcommand{\>}{\rangle}
\newcommand{\<}{\langle}
\renewcommand{\(}{\left(}
\renewcommand{\)}{\right)}

\renewcommand{\ll}{\left|}
\newcommand{\rl}{\right|}

\newcommand{\stf}{\,\,\,}

\newcommand{\stb}{\!\!\!}

\renewcommand{\k}{|}

\newcommand{\oro}{\hat{\rho}}
\newcommand{\oh}{\hat{h}}

\renewcommand{\d}{{\mbox d}}

\newcommand{\hb}{\hbar}

\renewcommand{\deg}{{^\circ}}

\begin{document}

\title{Quantum calculation of Coulomb reorientation and near-barrier fusion}


\author{C. Simenel}{
  address={DSM/DAPNIA, CEA SACLAY, F-91191 Gif-sur-Yvette, France}
  ,altaddress={NSCL, MSU, East
Lansing, Michigan 48824, USA}
}

\author{M. Bender}{
  address={DSM/DAPNIA, CEA SACLAY, F-91191 Gif-sur-Yvette, France}
  ,altaddress={NSCL, MSU, East
Lansing, Michigan 48824, USA}
}

\author{Ph. Chomaz}{
  address={GANIL, B.P. 55027,
F-14076 CAEN Cedex 5, France}
}

\author{T. Duguet}{
  address={NSCL, MSU, East
Lansing, Michigan 48824, USA}
  ,altaddress={Physics and
Astronomy Department, MSU, East Lansing, Michigan 48824, USA}
}

\author{G. de France}{
  address={GANIL, B.P. 55027,
F-14076 CAEN Cedex 5, France}
}

\begin{abstract}
We investigate the role of deformation on the fusion
probability around the barrier using the Time-Dependent
Hartree-Fock theory with a full Skyrme force. We obtain a
distribution of fusion probabilities around the nominal barrier
due to the different contributions of the various orientations of
the deformed nucleus at the touching point. It is also shown that
the long range Coulomb reorientation reduces the fusion
probability around the barrier.

\end{abstract}

\maketitle

\section{Introduction}

Fusion of massive nuclei has recently drawn a lot of
interest, especially at energies around the fusion barrier
generated by the competition between the Coulomb and nuclear
interactions. In this energy domain, the reaction mechanisms may
depend strongly on the structure of the collision partners. 
The proper description of near-barrier fusion is thus a
challenging N-body quantum dynamical problem involving the
competition between various reaction channels. For example,
the coupling between the internal degrees of freedom and the relative
motion may generate a fusion barrier distribution \cite{das83}.
Such couplings are needed to reproduce
 the  sub-barrier fusion \cite{das98}.

One of the internal degrees of freedom which can strongly affect the fusion
is the static deformation \cite{rie70,jen70}.
First, the fusion probability depends on the orientation of the deformed
nucleus at the touching point. Second, a reorientation
 can occur under the torque produced by the long-range Coulomb force
\cite{hol69,wil67,sim04,uma06b}. Such  a reorientation is a
consequence of the excitation of rotational states. It induces an
anisotropy in the orientation distribution, thus modifying the
near-barrier fusion  \cite{bab00}.

In this work 
we study the fusion of a
spherical and a prolate deformed nucleus 
%we study the fusion involving a prolate nucleus 
within the Time-Dependent Hartree-Fock (TDHF)
theory. We first show the effect of the orientation at the touching point on the
fusion probability. Then we include the long range Coulomb excitation
of rotational states
and study its effect on fusion.
The results give a useful interpretation of full coupling channels calculations.

\section{Time-Dependent Hartree-Fock theory}

Let us first recall briefly some aspects of  TDHF theory and
 of its numerical applications to nuclear collisions.
TDHF is a mean field quantum dynamical theory
\cite{har28,foc30,dir30} . It describes the evolution of 
occupied single particle wave functions
 in the mean field generated by all the particles.
The total wave function of the system is constrained to be
a Slater determinant at any time which assures an
exact treatment of the Pauli principle  during the dynamics.
All standard applications of TDHF neglect
pairing correlations so far. Like all mean-field methods, TDHF is best 
suited to desctibe average values of one-body operators.
%The pairing is not yet included in TDHF.
%It is optimized for the prediction of the average values of one
%body observables.
Such quantities are determined from the one-body density matrix
$
\oro=\sum_{n=1}^{N}\ll \az_{n}\> \<\az_{n}\rl
$
where $\k \az_n \>$ denotes an occupied single particle state.
In TDHF, its evolution is determined by a Liouville-von Neumann equation,
$
i\hb \partial_t \oro =[\oh(\ro ),\oro ]
$
where $\oh(\ro )$ is the mean-field Hamiltonian.

The great advantage of TDHF is that it treats the static
properties {\it and} the dynamics of nuclei within the same
formalism, i.e. using the same effective interaction (usually of
the Skyrme type \cite{sky56}). The initial state is obtained
through static Hartree-Fock (HF) calculations which are known to
reproduce rather well  nuclear binding energies and
deformations. TDHF can be used in two ways to describe nuclear
reactions:
\begin{itemize}
\item A single nucleus is evolved in an external field \cite{vau72},
simulating for instance the Coulomb field of the collision partner \cite{sim04}.
\item The evolution of two nuclei, initially with a zero overlap,
is treated in the same box with a single Slater determinant \cite{bon76,neg82}.
\end{itemize}
The first case 
is well suited for the description of inelastic scattering, like Coulomb 
excitation of vibrational and
rotational states. The second case is used for more violent
collisions like deep-inelastic and fusion reactions. In such
cases, the lack of a collision term in TDHF might be a drawback.
At low energy, however, the fusion is mainly driven by the one-body
dissipation because the Pauli blocking prevents 
nucleon-nucleon collisions.  The system fuses mainly by
transfering relative motion  into internal excitation via
one-body mechanisms well treated by TDHF.

Another important advantage of TDHF concerning its application
 to near-barrier reaction studies is that it contains
implicitely all  types of couplings between the relative
motion and internal degrees of freedom whereas in coupling
channels calculations one has to include them explicitely
according to physical intuition which is not always
straightforward for complex mechanisms. The  only condition in
TDHF is that the symmetries corresponding to the internal degrees
of freedom of interest are relaxed. This is now the case with the
latest TDHF codes in 3 dimensions (3D) which use a full Skyrme force
\cite{kim97,uma06a}. However, TDHF gives only classical
trajectories for the time-evolution  and
 expectation values of one-body observables. In particular,
TDHF does not include tunneling of the global wave function.

We use the TDHF code built by P. Bonche and
coworkers \cite{kim97}  using a Skyrme functional
\cite{sky56}. This code computes the evolution of each occupied
single-particle wave function in a 3D box assuming one symmetry plane.
 The step size of the network is 0.8~fm and the step
time 0.45~fm/c. We use the SLy4$d$ parametrization \cite{kim97} of
the Skyrme force  which is a variant of the SLy4 one
specifically designed for TDHF calculations.

\section{Fusion with a deformed nucleus}

\subsection{Effect of the static deformation}

Many nuclei exhibit static deformation, that is well described by 
mean-field calculations. Static deformation breaks the rotational invariance 
of the Slater determinant, which introduces an intrinsic frame of the nucleus.
TDHF calculations of nuclear collisions, however, are performed in the
laboratory frame, and one is left with an ambiguity concertning the relative 
orientation of the deformed nuclei. This is a critical point, because 
different orientations might ultimately lead to different reaction paths.

\begin{figure}[htbp]
  \includegraphics[height=.62\textheight]{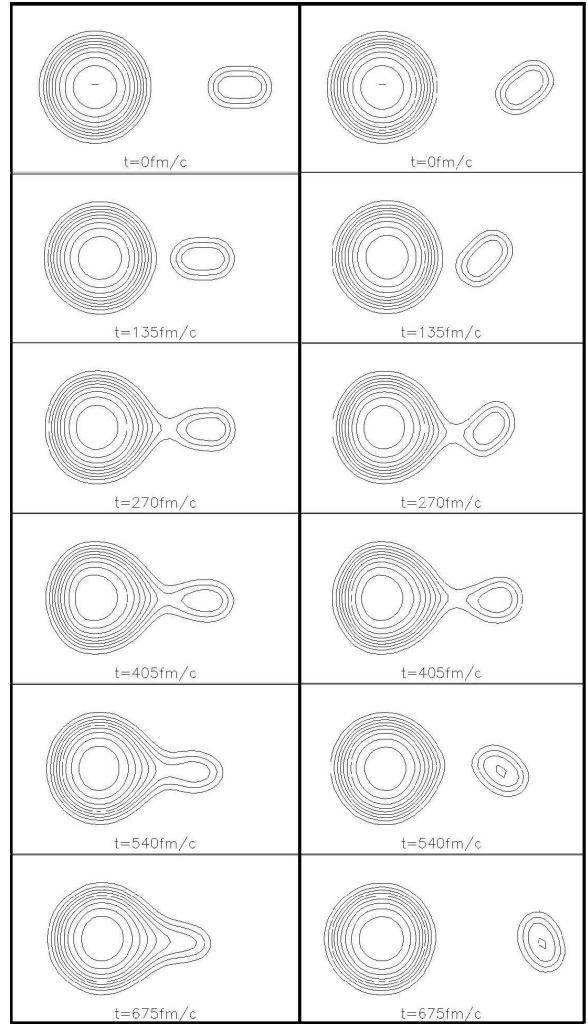}
  \caption{Density plots of head-on $^{208}$Pb+$^{24}$Mg collisions
  at $E_{CM}=95$~MeV with an initial orientation at 20~fm of 0$^\circ$ (left)
  and 37.5$^\circ$ (right). The time step between each figure is 135~fm/c.}
\label{fig:Tall}
\end{figure}

To illustrate this point
we consider central collisions of a prolate deformed $^{24}$Mg ($\beta_2=0.4$)
with a spherical $^{208}$Pb. For symmetry reasons,
the reaction mechanism will depend only on the energy
and the angle between the deformation axis and the collision axis
noted $\az$.
Fig. \ref{fig:Tall} shows the time evolution of the density
for two different initial orientations.
We see that with an initial orientation $\az = 0 \deg$ the nuclei fuse
 whereas with $\az = 37.5 \deg$ the two fragments
separate after a deep-inelastic collision.

The technic used to overcome the ambiguity of the initial orientation
 is based on two prescriptions \cite{sim04,uma06b}:
\begin{enumerate}
\item It is necessary to assume an initial distribution of orientations.
\item Interferences between different orientations are neglected.
Then each Slater determinant evolves
in its own mean field.
\end{enumerate}

\begin{figure}[htbp]
  \includegraphics[height=.2\textheight]{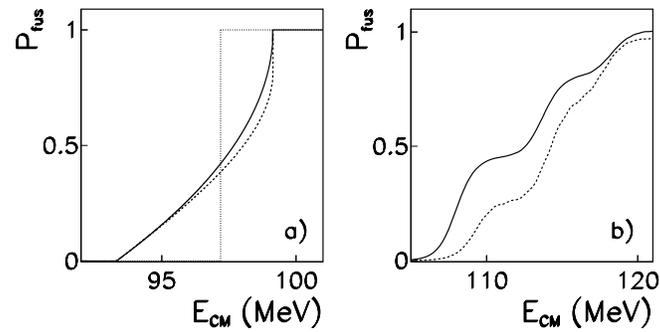}
  \caption{Fusion probability for a head-on collision or "penetrability"
  of $^{208}$Pb+$^{24}$Mg as a function
  of the c.m. energy. $a)$ TDHF results.
  Isotropic distribution of the initial orientations
  is supposed at 20~fm (solid line) and at 220~fm (dashed line).
  Step function expected in case of a spherical $^{24}$Mg (dotted line).
  $b)$ CCFULL results without (solid line) and with (dashed line) Coulomb excitation.}
\label{fig:pfus_e}
\end{figure}

Let us first assume an isotropic distribution of the orientations
at the initial time, corresponding to a distance $D=20$~fm
between the two centers of mass.
This means that the $^{24}$Mg is supposed to be initially in its
$0^+$ ground state and that all kind of
long range Coulomb excitations are neglected up to this distance.
Then, using the above prescriptions we get the fusion probability
$$
P_{fus}(E)=\frac{1}{2}\int_0^\pi \stb \d \az \stf \sin \az \stf P_{fus}(E,\az)
$$
where $P_{fus}(E,\az)=0$ or 1.
The solid line in Fig. \ref{fig:pfus_e}-a shows
the resulting fusion probability as function of the
center of mass energy. Below 93~MeV no orientation leads to fusion and
above 99~MeV, all of them fuse.
Between these two values, the higher the energy,
the more orientations lead to fusion.
As shown in Fig. \ref{fig:Tall},
configurations with small $\az$ are the first to fuse,
 even below the nominal barrier which would correspond
to a spherical $^{24}$Mg case (dotted line).
To conclude, sub-barrier fusion is described in TDHF through
couplings between static deformation and relative motion.

\subsection{Long range Coulomb reorientation}

As a consequence of numerical limitations, actual TDHF calculations for 
collisions are performed in relatively small boxes and are started with
internuclear distances of a few Fermi. However, the Coulomb interaction
starts playing a role much earlier in the reaction process 
\cite{alder}. It is weaker than the nuclear
interaction, but integrated over a long time it may induce a
polarization, and then modify the reaction mechanism
\cite{sim04,row06}. Long range Coulomb reorientation has been
studied in Ref. \cite{sim04} with TDHF. The results have been
interpreted using the classical formalism \cite{alder,broglia}
where the motion of a deformed rigid projectile is considered in
the Coulomb field of the target. An important conclusion of this
work is that the reorientation, although being the result of a
Coulomb effect, depends neither on the charges nor on the energy.
Let us illustrate this phenomenon with a simple example. Consider
a system at time $t$  formed by a deformed projectile at the
distance $D(t)$ from the target. Increasing artificially the
charge of one of the nuclei at this time has two effects. First,
the Coulomb interaction increases and then the torque applied on
the deformed nucleus should {\it increase} too.  On the other
hand, the distance $D(t)$ between the projectile and the target is
larger because of the stronger Coulomb repulsion between the
centers of mass. The latter effect leads to a {\it decrease} of
the effective torque at time $t$ and both effects overall cancel
exactly. One is left with a charge independent reorientation. The
same argument applies for the energy.

 To study the effect of reorientation on fusion 
 we calculate the reorientation in the approach phase
between $D=220$~fm and 20~fm with TDHF using the technic described
in Ref. \cite{sim04}. Assuming an isotropic distribution of orientations
at 220~fm we get a new distribution at 20~fm which includes the reorientation
coming from long range Coulomb excitation.
The new fusion probability distribution (dashed line
in Fig. \ref{fig:pfus_e}-a) is obtained with two additional assumptions:
\begin{itemize}
\item The rotational speed of the $^{24}$Mg is neglected
at the initial time of the TDHF calculation (corresponding to
$D=20$~fm), i.e. only a static reorientation is considered.
\item The effect of the excitation energy on the relative motion is neglected,
i.e. we assume a Rutherford trajectory before $D=20$~fm.
\end{itemize}
We observe in Fig. \ref{fig:pfus_e}-a a fusion hindrance up to 20$\%$
which is due to higher weights on orientations
leading to compact configurations ($\az \sim 90\deg$ at the touching point)
because of the reorientation \cite{sim04,row06}.

The previous study is helpful to interpret coupling channels
results. Calculations on the same system have been performed with
the code CCFULL \cite{hag99} including coupling to the five first
excited states of  $^{24}$Mg rotational band. The
fusion probability, or "penetrability" of the fusion barrier is
given by the relation $ P_{fus}=\frac{\d\(\si E\)}{\d E}\pi
R_B^2 $ where $R_B=11.49$~fm is the barrier position. Fig.
\ref{fig:pfus_e}-b shows the fusion probability obtained from
CCFULL including nuclear (solid line) and nuclear+Coulomb (dashed
line) couplings. As with TDHF, an hindrance of the fusion due to
Coulomb couplings is observed. However the shape of TDHF and
CCFULL distributions are quite different. This is due to the fact
that quantum mechanical effects are missing in TDHF. This point
out the importance of improving the theory. It is also striking to
see that TDHF "misses" the nominal barrier by about 15$\%$. TDHF is
known to overestimate the fusion cross sections. One possible
issue might be the time odd terms in the Skyrme energy functional.
Their importance on fusion have been stressed recently
\cite{uma06a,mar06}.

\section{conclusion}

To summarize, we performed a TDHF study of near-barrier fusion 
between a spherical and a deformed nucleus.
The calculations show that, around the barrier,
different orientations lead to different reaction path.
Considering all possible orientations leads to a
distribution of fusion probabilities interpreted as an effect
of the coupling between the static deformation and the relative motion.
We then included the long range Coulomb coupling which induces
a {\it charge} and {\it energy} independent reorientation
of the deformed nucleus. The effect of the reorientation is
to hinder the near-barrier fusion.
Finally the TDHF study have been used to interpret
coupling channels calculations with the code CCFULL which show
also an hindrance of near-barrier fusion due to Coulomb couplings.
We also note some drawbacks of TDHF which, in one hand,
underestimates the fusion barrier, and, in the other hand,
miss important quantum effects.

\begin{theacknowledgments}
We warmly thank Paul Bonche for providing his TDHF code.
This work has been partially supported by NSCL, Michigan State University
and the National Science Foundation under the grant PHY-0456903.
\end{theacknowledgments}

\bibliographystyle{aipproc}

\end{document}